\documentclass[twocolumn,preprintnumbers,prb]{revtex4}
\usepackage{graphicx}
\usepackage{amsmath}
\usepackage{amssymb}
\usepackage{chngcntr}
\newcommand{\bfm}[1]{\textbf{\em#1}}

\begin{document}

\title{X-ray dynamical diffraction in amino acid crystals: a step towards improving structural resolution of biological molecules via physical phase measurements}

\author{S\'ergio L. Morelh\~ao}
\email[corresponding author: ]{morelhao@if.usp.br}
\affiliation{Institute of Physics, University of S\~ao Paulo, S\~ao Paulo, SP, Brazil}
\author{Cl\'audio M. R. Rem\'edios}
\affiliation{Faculdade de F\'{i}sica, Universidade Federal do Par\'a, Bel\'em, PA, Brazil}
\author{Guilherme Calligaris}
\affiliation{Instituto de F\'{i}sica, Universidade Estadual de Campinas, Campinas, SP, Brazil}
\author{Gareth Nisbet}
\affiliation{Diamond Light Source, Harwell Science and Innovation Campus, OX11 0DE, UK}

\date{\today}

\begin{abstract}
In this work, experimental and data analysis procedures were developed and applied for studying amino acid crystals by means of X-ray phase measurements. It clearly demonstrated the sensitivity of invariant triplet phases to electronic charge distribution in D-alanine crystal, providing useful information for molecular dynamics studies of intermolecular forces. Feasibility of phase measurements to investigate radiation damage mechanisms is also discussed on experimental and theoretical grounds.
\end{abstract}

\maketitle

\section{Introduction}

Hydrogen bond is the most important of all directional intermolecular interactions, it is ubiquitous in nature and a critical chemical bond in life science responsible for the conformational stability of proteins ensuring their biological functionality \cite{ste02,ros15}. Within the current context of experimental and theoretical methods for molecular structure determination there are still many challenges and, among them, accurate description of interactions between an electron-deficient hydrogen atom and electron-rich atoms \cite{rei12,taf16}. Particulary in protein X-ray crystallography, the detection of H atoms is one of the major problems, since they display only weak contributions to diffraction data \cite{hid15}. Nuclear methods such as neutron diffraction are sensitive to the proton position, and combined with X-ray methods have been able to locate important H atoms to improve our understanding of macromolecular structure and function \cite{mpb15}. However, even in small molecule crystals, experimental determination of electron charge in hydrogen bonds is a difficult problem, demanding charge density maps with sub-\r{a}nstr\"om resolution \cite{rsg00,kra12}.

Radiation damange in X-ray crystallography is another problem that compromises the resolution of electron density maps as well as the reliability of structure determination in biomolecules and organic samples in the crystalline state \cite{ten00,mpb15,ger15,gar17}. Despite all the advances in X-ray detectors and data collection protocols, radiation damage still occurs at cryogenic temperatures and the known protein structures suffer, at least to some extent, from inaccuracies originating from this effect \cite{poz13}. Formation of hydrogen gas in the sample during irradiation, rather than bond cleavage, has been pointed out as the major cause for the loss of high-resolution information \cite{am10}. The largely incomplete understanding of the physical and chemical mechanisms behind structural damage have recently motivated the development of computational tools specifically for investigating damage creation mechanisms \cite{lb16}. In this sense, it is desirable to have an X-ray tool capable of probing experimentally small structural features such as electron charge in hydrogen bonds, radiation damage effects at atomic scales, or simply to validate high resolution structures obtained from other experimental or purely computational methods.

\subsection{Physical phase measurements in X-ray crystallography}

From inorganic crystals to protein crystals, structure determination with atomic resolution is mostly based on diffraction techniques (X-rays, neutron, and electrons). However, since the coherent scattering cross-section for X-ray by atoms have intermediate values between those for electrons and neutrons, physical measurements of structure factor phases have been feasible with X-rays only \cite{zga14}. Dynamical diffraction taking place within perfect domains is another requirement for physical phase measurements via multiple diffraction (MD) experiments. In crystals with small unit cells, dynamical diffraction regime is achieved in much smaller domains than in crystals with large cells such as protein crystals. Fact that has allowed phase measurements to reveal structural details---inaccessible by other techniques---in magnetic materials \cite{she06}, optical crystals with dopant ions \cite{slm11,zga14}, and to solve chirality in crystals with no resonant atoms \cite{kh95,qs00,slm15}.

\begin{figure*}
\includegraphics[width=5.4in]{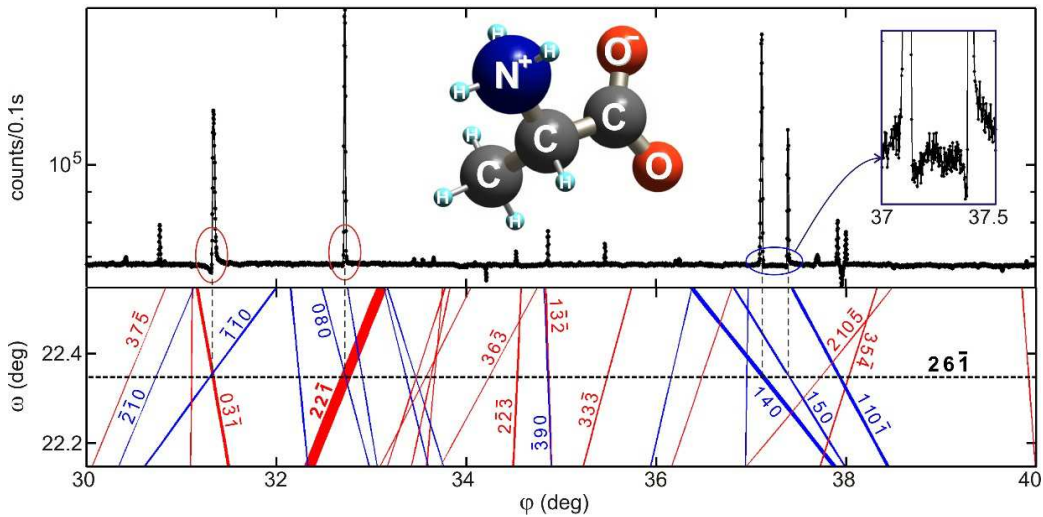}\\
\caption{Dynamical diffraction in D-alanine crystal giving rise to asymmetric intensity profiles in MD cases (top panel). X-rays of 10\,keV, $\sigma$ polarization. Inset: D-alanine zwitterion. Graphical indexing through Bragg-cone lines (bottom panel) provides a general picture of the nearby cases, their relative strength (line thickness), and easy distinction of the out-in/in-out geometries (blue/red lines). MD peaks are seen at the intersections of Bragg-cone lines with the $26\bar{1}$ one (horizontal dashed line).}
\label{fig:xrs} \end{figure*}

Excitation of 2nd-order diffractions, MDs for short, and their potential applications in X-ray crystallography have been investigated since \onlinecite{ren37} performed the first azimuthal scanning in the early 20th century, the so called Renninger scanning. Every time similar experiments are carried out, very often the intensity profiles exhibit characteristic asymmetries, such as those seen in Fig.~\ref{fig:xrs} (top panel), owing to dynamical coupling of the simultaneously diffracted waves inside a single crystal domain. Over several decades, these often observed asymmetries have motivated numerous researchers in developing theoretical approaches and experimental procedures to process MD intensity profiles into structural information \cite{har61,col74,pos77,chp81,jur82,slc97,wec97,slc99,cmw01,fmo02,slm02,she03}.

Nowadays, crystallographic studies are conducted by an increasing number of non-experts due to substantial instrumental automation and the continuing improvement of softwares \cite{poz13}. In this scenario, old phase measurement methods based on dynamical diffraction simulation to obtain triplet phase values within error bars are completely outdated, so that the average number of publications using this technique has dropped to less than one per year after the mid-2000s. Besides time consuming experiments, familiarity with dynamical theory, and high level of instrumental expertise in single crystal diffraction, the major reason discouraging further exploitation of the technique was indeed the low accuracy of the obtained phase values providing no gain in structural resolution \cite{soa03}. However, it is well known for quite a long time that the type of asymmetry, i.e. if the MD intensity profile has lower/higher (L$|$H) or higher/lower (H$|$L) shoulders, is a reliable information even in crystals with some mosaicity \cite{slc84,she86,wec97,tho03,slm03}. Very recently, it has been proposed that this fact leads to a window of accuracy in phase measurements, implying in new strategies on how to look at these asymmetries, and opening opportunities for high resolution studies of crystal structures \cite{slm15}.

In this work, to demonstrate in practice one of such strategies and to highlight its potential in structural biology, we choose the challenge of detecting electron charge in hydrogen bonds responsible for intermolecular forces between amino acid molecules. The strategy is described step-by-step from experiment planning to data analysis procedures. Easy computer codes are used and no dynamical diffraction simulation is needed. Reliable phase information are identified by a simple graphical indexing, e.g. Fig.~\ref{fig:xrs} (bottom panel), which is also very useful for other diffraction techniques in semiconductor devices and single crystals in general \cite{jzd16,gan15}. Diffraction data in single crystals of D-alanine collected at two synchrotron facilities and with different instrumentation (flux, optics, and goniometry) are presented. Model structures taking into account ionic charges are proposed and refined throught comparison with experimental data, leading to an ideal model to describe X-ray diffraction by this simple amino acid molecule in terms of triplet phase invariants. According to this model, van der Waals forces between D-alanine zwitterions are also acting in the crystal structure. Moreover, within our data set, we found the first insight on the possibility of using X-ray phase measurements to study radiation damage in crystals.

\section{Model structures}

\begin{figure}
\includegraphics[width=2.7in]{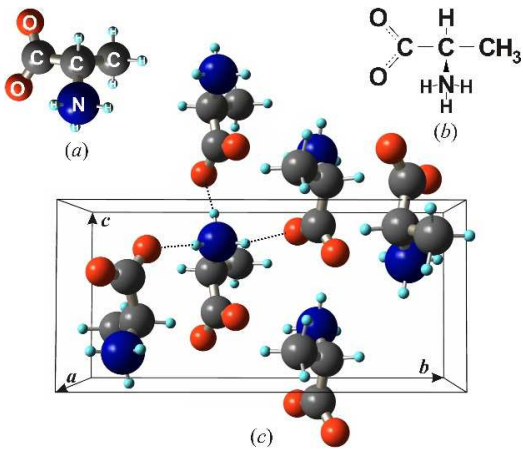}\\
\caption{(a,b) D-alanine molecule, 3D and plain view. (c) N--H$\cdots$O bonds (dashed lines) between adjacent molecules in the crystal structure, orthorhombic unit cell of lattice parameters $a=6.031(3)$\,\AA, $b=12.335(5)$\,\AA, and $c=5.781(3)$\,\AA. 12 N--H$\cdots$O bonds per unit cell. }
\label{fig:vdWaals}
\end{figure}

With molecular formula C$_3$H$_7$NO$_2$, L- and D-alanine are the smallest molecule among the amino acids. When grown in aqueous solution, both enantiomers crystallize in space group $P2_12_12_1$ at ambient pressure, with four molecules per unit cell. Intermolecular forces are hydrogen bonds where the amine group (NH$^{3+}$) of each molecule, in its zwitterionic form \cite{evb07,jwm11}, makes N--H$\cdots$O bonds with oxygen atoms of three carboxylate groups (COO$^-$) of nearest molecules, Fig.~\ref{fig:vdWaals}, thus linking the molecules together to form a three-dimensional crystal structure \cite{deg08,npf10}. Due to these hydrogen bonds there is a non-spherosymmetric electron charge distribution around each amine group.

For successful use of phase measurements, the first and fundamental step in any application of this technique is the identification of MD cases suceptible to the specific structural features under investigation. It is acomplished by elaborating suitable model structures for each particular study. In our example here, we are seaching for MD cases suceptible to the non-spherosymmetric electron charge distribution due to hydrogen bonds, and for this goal two simple models are initial used. One is a realistic model, tagged as the NH3 model, where the hydrogen atoms are set around the nitrogens at distances of $1.05\pm0.02$\,\AA, Fig.~\ref{fig:vdWaals}(a), as determined by neutron diffraction \cite{msl72,ccw05}. The other is a hypothetical model, tagged as the N$3e$ model, where hydrogen electrons are placed in the nitrogen orbitals so that the amine group scatter X-rays as the N$^{3-}$ ion with spherosymmetric charge distribution.

\begin{figure*}
\includegraphics[width=5.4in]{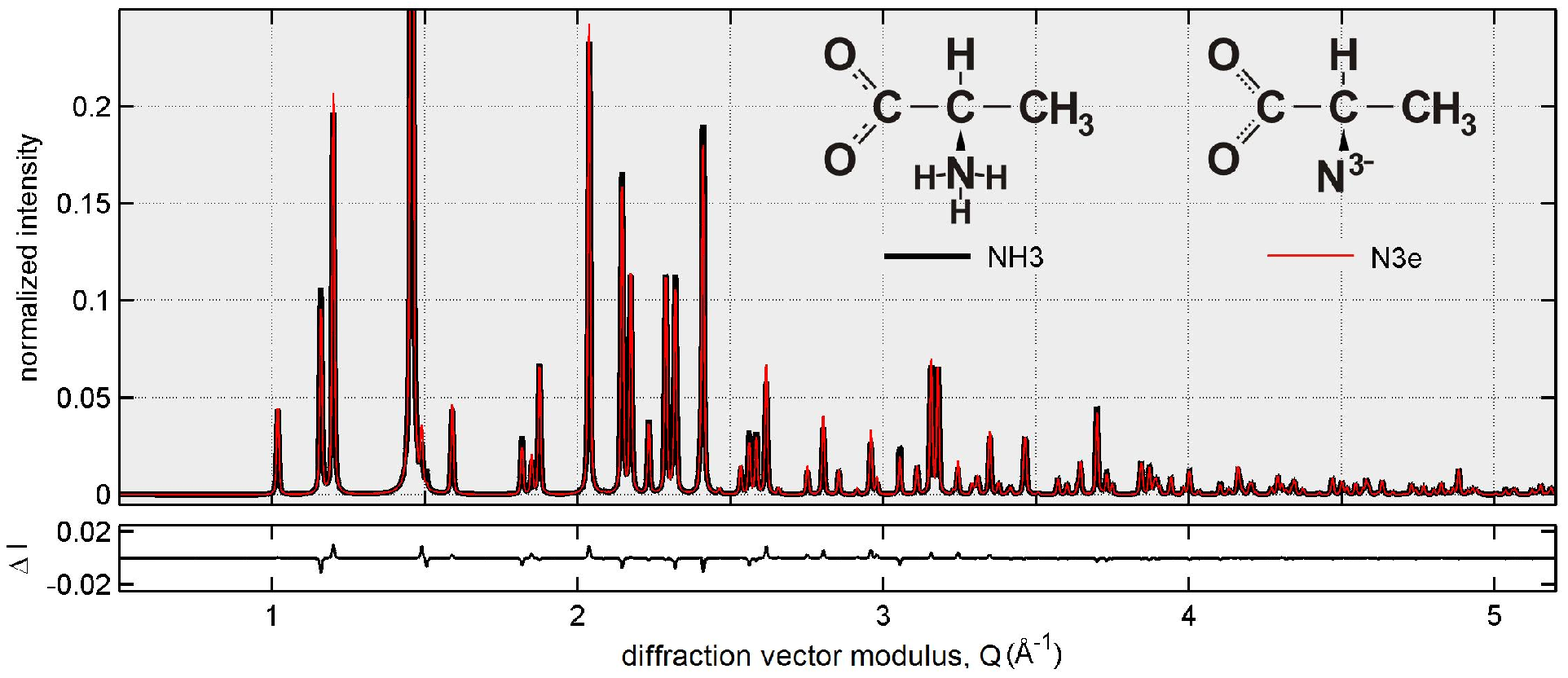}\\
  \caption{Comparison of simulated XRD patterns according to NH3 and N$3e$ model structures. X-rays of 10\,keV, $\sigma$ polarization.} \label{fig:xrdcomparison}
\end{figure*}

In terms of diffracted intensities, the overall differences can be seen by comparing simulated X-ray powder diffraction patterns for both models in Fig.~\ref{fig:xrdcomparison}. Tabulated atomic scattering factors for neutral atoms \cite{ep06} were used in calculating diffracted intensities of the NH3 model structure, while the atomic scattering factor of the N$^{3-}$ ion \cite{slm15} stands for the whole scattering of amine groups in the N$3e$ model. The comparison in Fig.~\ref{fig:xrdcomparison} shows that to distinguish between these models by such standard X-ray method, experimental accuracy better than 1\% (regarding the main peak) in measuring relative intensities of diffraction peaks would be required. That is the reason by which the realistic model NH3 had to be based on neutron diffraction data, but where no information is available on the polarization state of H atoms.

\section{Principles of phase measurements}

Phase measurements rely on the fact that in a crystal undergoing dynamical diffraction, the integrated intensity of one reflection, reflection $G$, when measured as a function of the excitement of another reflection, reflection $H$, gives rise to an intensity profile whose asymmetry depends on the triplet phase
\begin{equation}\label{eq:psi}
    \Psi = \delta_H + \delta_{G-H}-\delta_G\,,
\end{equation}
e.g. \onlinecite{slc97}, where $\delta_X$ is the phase of structure factor $F_X$ of reflection $X$ ($X = G$, $H$, and $G-H$).

\begin{figure}
\includegraphics[width=2.7in]{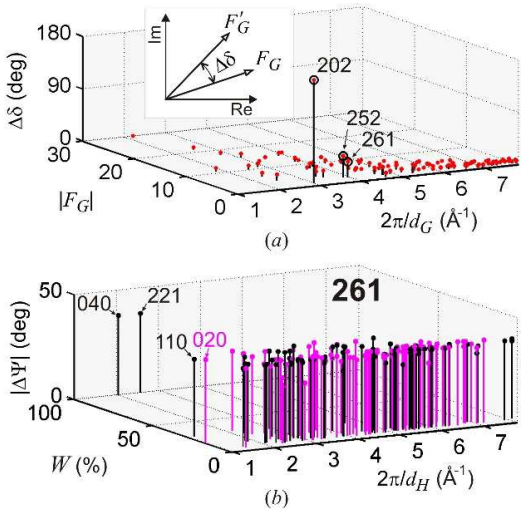}\\
  \caption{(a) Difference $\Delta\delta$ in structure factor phases regarding the proposed models, as detailed in the inset (NH3\,$\rightarrow F_G$ and N$3e\rightarrow F^\prime_G$). X-rays of 10\,keV. (b) Three-beam cases predicted to show opposite profile asymmetries on each model structure according to the criteria $\cos(\Psi)\cos(\Psi^\prime)<1$. $G=261$ and $\Delta\Psi=\Psi^\prime-\Psi$ (NH3\,$\rightarrow\Psi$ and N$3e\rightarrow\Psi^\prime$). Limited to amplitude $W>5$\,\%. $d_{G,H}$ is the interplanar distance of Bragg planes. } \label{fig:comparators}
\end{figure}

To identify the most susceptible MD cases for studying hydrogen bonds in this amino acid crystal by phase measurements, it is necessary first to search for structure factors with phases susceptible to changes in the models, as done in Fig.~\ref{fig:comparators}(a). It indicates a few reflections, such as 202, 252, and 261, that are good candidates for phase measurements. Since these reflections have small $|F_G|$ values, i.e. are weak reflections, they can only be used as the primary $G$ reflection. After selecting the $G$ reflection, it is necessary to figure out secondary $H$ reflections that promote MD cases with opposite profile asymmetries for each proposed model structures. It can be done by calculating
\begin{equation}\label{eq:W}
    F_H F_{G-H}/F_G = W{\rm e}^{i\Psi}
\end{equation}
for both models and picking up the cases where the phase shift $\Delta\Psi$ is large enough to make the triplet phase to go across the $\pm90^\circ$ values, i.e. those cases where $\cos(\Psi)\cos(\Psi+\Delta\Psi)<1$. This procedure is illustrated in Fig.~\ref{fig:comparators}(b) for 261 as the $G$ reflection. It foresees many cases having opposite asymmetries, including the cases for the 221 and 040 secondary reflections with the largest relative values of the amplitude $W$ (see a partial list in Table~\ref{tab:wlist1}).

\begin{figure}
\includegraphics[width=2.7in]{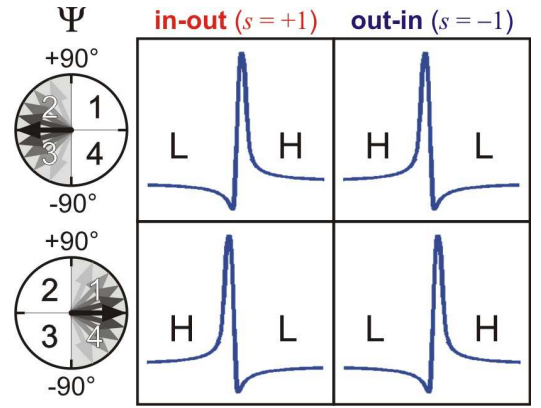}\\
  \caption{Criteria of L$|$H ($s\cos(\Psi)<1$) and H$|$L ($s\cos(\Psi)>1$) or  profile asymmetry in three-beam diffraction with respect to the in-out ($s=+1$) or out-in ($s=-1$) geometry of excitement and interval of values of the triplet phase $\Psi$: $\cos(\Psi)<1$, quadrants 2 and 3 (top panels), or $\cos(\Psi)>1$, quadrants 1 and 4 (bottom panels). } \label{fig:asycrit}
\end{figure}

\section{Graphical indexing of Renninger scans}

With a list of susceptible phases in hands, another very important step is to have an efficient method to select the most easy-to-measure MD cases capable of providing reliable phase information. A graphical indexing method based on two-dimensional representation of Bragg cones (BC) is used here for sake of clarity in the data analysis, \S\ref{ss:msref} and \S\ref{ss:rdamage}. For any reflection of diffraction vector ${\bfm Q}$, its two-dimensional BC  representation is given by the relationship
\begin{equation}\label{eq:bcl}
    \cos(\varphi-\alpha_Q) = \frac{\sin\theta-\sin\omega\,\cos\gamma_Q} {\cos\omega\,\sin\gamma_Q}\;.
\end{equation}
$\omega$ and $\varphi$ are the instrumental angles describing the incident wavevector $${\bfm k} = -\frac{2\pi}{\lambda}[\cos\omega\,\cos\varphi,\, \cos\omega\,\sin\varphi,\, \sin\omega]$$ on a sample $xyz$ frame where $z$ is along the azimuthal rotation axis \cite{jzd16}. The $\alpha_Q$ and $\gamma_Q$ angles are obtained by projecting the diffraction vector on this crystal frame, i.e., $${\bfm Q}=|{\bfm Q}|\,[\sin\gamma_Q\,\cos\alpha_Q,\,\sin\gamma_Q\,\sin\alpha_Q,\,\cos\gamma_Q]\,,$$ $|{\bfm Q}|=(4\pi/\lambda)\sin\theta$, and ${\bfm k}\cdot{\bfm Q}=-|{\bfm Q}^2|/2$.

Eq.~(\ref{eq:bcl}) provides two solutions for the azimuth $\varphi$ as a function of the incidence angle $\omega$. These solutions represent the two possible excitation geometries that are plotted as the out-in (blue) and in-out (red) BC lines in the $\omega\!:\!\varphi$ graphs, e.g. Fig.~\ref{fig:xrs}. Using lines of different colors to identify each one of these solutions is quite helpful since the observed profile asymmetries depend on both phase and excitation geometry, as summarized in Fig.~\ref{fig:asycrit}. Another useful trick for graphically indexing Renninger scans is plotting BC lines with relative thickness. Here we use line thicknesses proportional to the amplitude $W$, Eq.~(\ref{eq:W}). For instance, in the Renninger scan of reflection $26\bar{1}$ in Fig.~\ref{fig:xrs}, the strongest peak has the thickest BC lines due to secondary $22\bar{1}$ reflection.

\section{Experimental}

Single crystals of D-alanine were grown by slow evaporation from supersaturated aqueous solutions: D-alanine powder (98\% purity) diluted in distilled water, concentration of 0.25\,g/ml, and pH between 6 and 6.5. The solution was kept at a constant temperature (296K) in a beaker covered with perforated plastic lid during three weeks. Transparent single crystals showing well-formed natural faces were obtained, such as the one used here with approximated dimensions of $4\times3\times10\,{\rm mm}^3$ and largest face as the (130) planes (\S\ref{ap:azscan}, Fig.~\ref{fig:fullxrs}). Lattice parameters $a=6.031(3)$\,\AA, $b=12.335(5)$\,\AA, and $c=5.781(3)$\,\AA\, were determined by X-ray powder diffraction in another sample of the same batch, and they agree very well with the MD peak positions within an accuracy of $0.01^\circ$.

X-ray data acquisition were carried out at the Brazilian Synchrotron Light Laboratory (LNLS), beamline XRD2, and at the Diamond Light Source, beamline I16, testing advantages and limitations of two possible procedures for measuring line profiles of MD peaks. In one procedure, both adjustment arcs of the goniometric head are used to physically align the primary diffraction vector with one rotation axis of the sample stage. Wide azimuthal scans are possible, although eventual corrections of the incidence angle are necessary depending on the residual misalignment between the diffraction vector and the rotation axis. This procedure is preferred in term of accuracy in both line profile and position of the peaks \cite{rof07}, although after fixing the sample it is limited to reflections that can be aligned within the $\pm20^\circ$ range of the adjustment arcs. Rotating crystal method for indexing and pre-alignment of accessible reflections have been used, as demonstrated in \S\ref{ap:cara}.

In the other procedure, the azimuthal scans are performed by combining rotations of the diffractometer axes. This multi-axis goniometry is the standard procedure in single crystal diffractometers. The sample is fixed at the holder within eye accuracy and after finding two non parallel reflections, the crystal orientation matrix is built. With a proper script for azimuthal scanning, it is possible to inspect many primary reflections automatically. But, the number of accessible MD cases and data accuracy depend on angular range of combined rotations and sphere of confusion of the used diffractometer.

Despite of the distinct instrumentations at the used beamlines, energy and angular resolution were nearly the same: spectral width of $2\times10^{-4}$ and beam divergences of 0.1\,mrad. Brightness of the beam at I16 have required some attention to avoid fast radiation damage to fragile crystals stabilized by hydrogen bonds such as D-alanine. Exposure to the direct beam cause immediate damage, e.g. the streaks seen at the (130) surface in Fig.~\ref{fig:fullxrs} (inset). Primary $26\bar{1}$ was measured with the physical alignment procedure and X-rays of 10\,keV, while the multi-axis goniometry was used to measure a few MD cases with primaries 261 and 080, and X-rays of 8\,keV. Only the primary 080 was measured in Laue-transmission geometry regarding the entrance surface (130), all others in Bragg-reflection geometry. Vertical scattering plane ($\sigma$ polarization) has been used in all measurements where the asymmetry criteria in Fig.~\ref{fig:asycrit} apply for most MD cases. For other polarizations these criteria must be reviewed \cite{yps00,slm01,slm02}.

\begin{table}
\caption{Theoretical triplet phases according to structure models NH$_3$ ($\Psi$) and N$3e$ ($\Psi^\prime$) of D-alanine for a few secondary $H$ reflections seen in Fig.~\ref{fig:xrs}, b/r letters stand for blue/red BC lines. Experimental peak asymmetries are given in terms of parameter $R_a$, Eq.~(\ref{eq:R}). Relative amplitudes and positions are estimated by $W$ and $\varphi_0$ values, respectively.}\label{tab:phases}
\begin{center}
\begin{tabular}{crrcrcc}
\hline\hline
  $H$ & $\Psi\,(^\circ)$ & $\Psi^\prime\,(^\circ)$ & $W$(\%) & \multicolumn{2}{c}{$R_a$} & $\varphi_0\,(^\circ)$ \\
\hline
$3 7 \bar{5}$r & $65$ & $104$ & 6 & 2.2 & (H$|$L) & $30.415$ \\
$\bar{2} \bar{1} 0$b & $-111$ & $-74$ & 9 & 2.2 & (H$|$L) & $30.773$ \\
$0 \bar{3} \bar{1}$r & $165$ & $-158$ & 29 & $-33.8$ & (L$|$H) & $31.341$ \\
$\bar{1} \bar{1} 0$b & $-3$ & $31$ & 16 & --- & --- & $31.341$ \\
$ 2 2 \bar{1}$r & $-98$ & $-60$ & 100 & $-12.6$ & (L$|$H) & $32.722$ \\
$ 0 8 0$b & $69$ & $106$ & 7 & --- & --- & $32.722$ \\
$ 2 \bar{2} \bar{3}$r & $73$ & $108$ & 11 & 1.0 & (H$|$L) & $34.527$ \\
$1 \bar{3} \bar{2}$r & $0$ & $31$ & 12 & 3.7 & (H$|$L) & $34.863$ \\
$\bar{3} 9 0$b & $-111$ & $-72$ & 6 & --- & --- & $34.863$ \\
$3 3 \bar{3}$r & $-116$ & $-72$ & 11 & $-1.1$ & (L$|$H) & $35.463$ \\
$1 4 0$b & $100$ & $143$ & 38 & 32.8 & (H$|$L) & $37.114$ \\
$1 5 0$b & $-3$ & $32$ & 16 & $-11.5$ & (L$|$H) & $37.391$ \\
\hline\hline
\end{tabular}
\end{center}
\end{table}

\section{Results and Discussions}

\begin{figure*}
\includegraphics[width=5.4in]{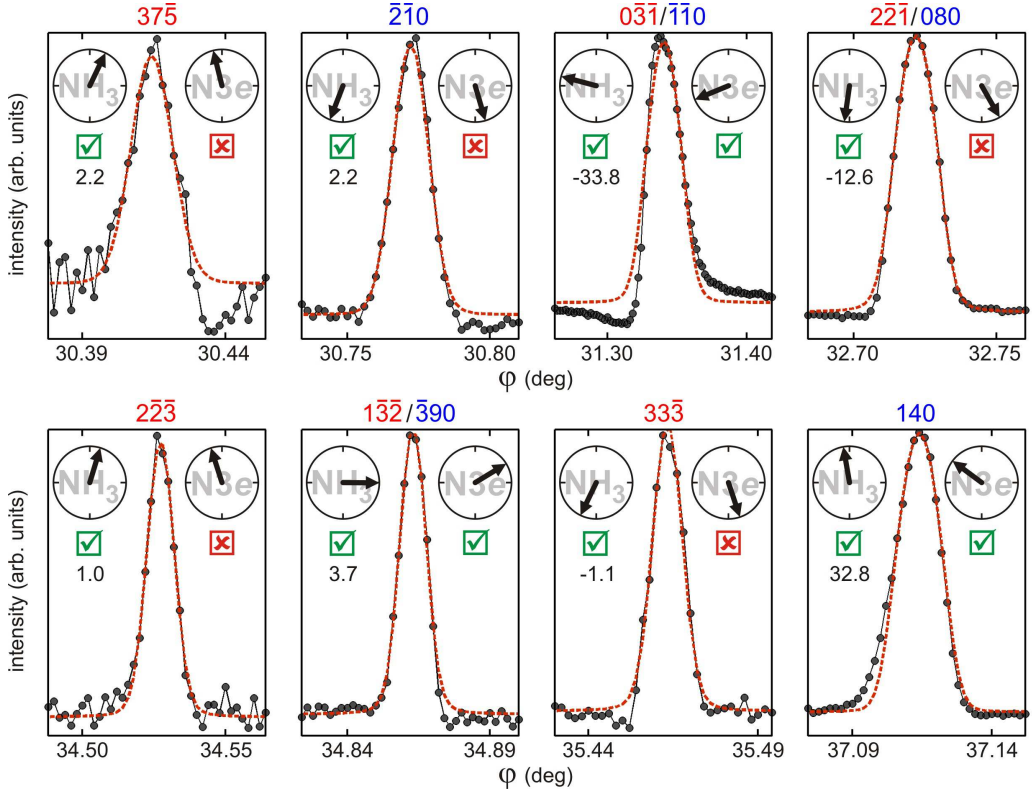}\\
\caption{Analysis of MD peak-profile asymmetry in D-alanine. Experimental profiles (closed circles connected by lines) from Fig.~\ref{fig:xrs} shown against data fitting with a symmetrical function (dashed-red lines). Primary reflection $G=26\bar{1}$. $H$ reflections (blue/red indexes for out-in/in-out geometries), triplet phase values for both model structures (arrows), and their compatibility (checkbox) with the observed profile asymmetries ($R_a$ values at left) are indicated for each peak, as well as in Table~\ref{tab:phases}.}
\label{fig:asyprofs}
\end{figure*}

Line profile asymmetries have been determined according to the value of
\begin{equation}\label{eq:R}
   R_a = 2\left(\Delta I_L-\Delta I_R\right)/\sqrt{\min \{I_s\}}
\end{equation}
where $$ \Delta I_L = \frac{1}{N_{\varphi_j<\varphi_0}} \sum_{\varphi_j<\varphi_0}[I_e(\varphi_j)-I_s(\varphi_j)]$$ and $$\Delta I_R = \frac{1}{N_{\varphi>\varphi_0}} \sum_{\varphi_j>\varphi_0}[I_e(\varphi_j)-I_s(\varphi_j)]\,.$$ $N_{\varphi_j<\varphi_0}$ and $N_{\varphi_j>\varphi_0}$ are the number of data points at the left and right side of the diffraction peak, respectively. $\Delta I_{L,R}$ stand for the mean intensity difference on each side of the peak since $I_e(\varphi_j)$ is the $j$th experimental data point and $I_s(\varphi_j)$ is the corresponding point obtained by data fitting with a symmetric pseudo-Voigh function, which also provides the peak center $\varphi_0$. The asymmetric character of each intensity profile is therefore given as H$|$L when $R_a\geq1$ or L$|$H when $R_a\leq-1$. Diffraction peaks are considered symmetric, i.e. with an indistinguishable type of asymmetry, when $|R_a|<1$. A few examples of data fitting by symmetric line profile function are shown in Fig.~\ref{fig:asyprofs}, and their corresponding triplet phase values are given in Table~\ref{tab:phases}.

Compatibility analysis between experimental asymmetries and proposed models is carried out on bases of a true/false test according to
\begin{equation}\label{eq:truefalse}
    s\cos(\Psi)R_a\left\{
                \begin{array}{ll}
                  >1\;\Rightarrow\;{\rm true}\\
                  <1\;\Rightarrow\;{\rm false}
                \end{array}
              \right.\;,
\end{equation}
which is reliable as far as $|R_a|\geq1$. The true/false outcomes for each model are indicated at checkboxes aside each experimental profile in Fig.~\ref{fig:asyprofs}. Even profile asymmetries in the symmetric/asymmetric limit where  $|R_a|\gtrsim 1$ can be also classified within eye accuracy, as those at $\varphi_0 = 34.53^\circ$ ($R_a=1$) and $35.46^\circ$ ($R_a=-1.1$). In either ways, all profile asymmetries are consistent (true) for the NH3 model only.

Let us emphasize what has been accomplished so far. By selecting just a few MD cases, Fig.~\ref{fig:asyprofs}, within a narrow Renninger scan of no more than 10$^\circ$, we already demonstrate experimentaly the existance of a non-spherosymmetric electron density due to H atoms around the amine group. It is an amazing result with respect to the current methods in crystallography where, to do similar demonstration, would be necessary to collect thousands of reflections and solve the phase problem for constructing high-resolution electron density maps of the amine group as in \onlinecite{rsg00}, or to combine diffraction data and calculations of periodic density functional theory as in \onlinecite{npf10}. However, more refined models than the NH3 are needed to explain intermolecular forces stabilizing the crystal structure, which would not exist if all atoms are neutral and unpolarized.

\subsection{Model structure refinement}\label{ss:msref}

\begin{figure}
\includegraphics[width=2.7in]{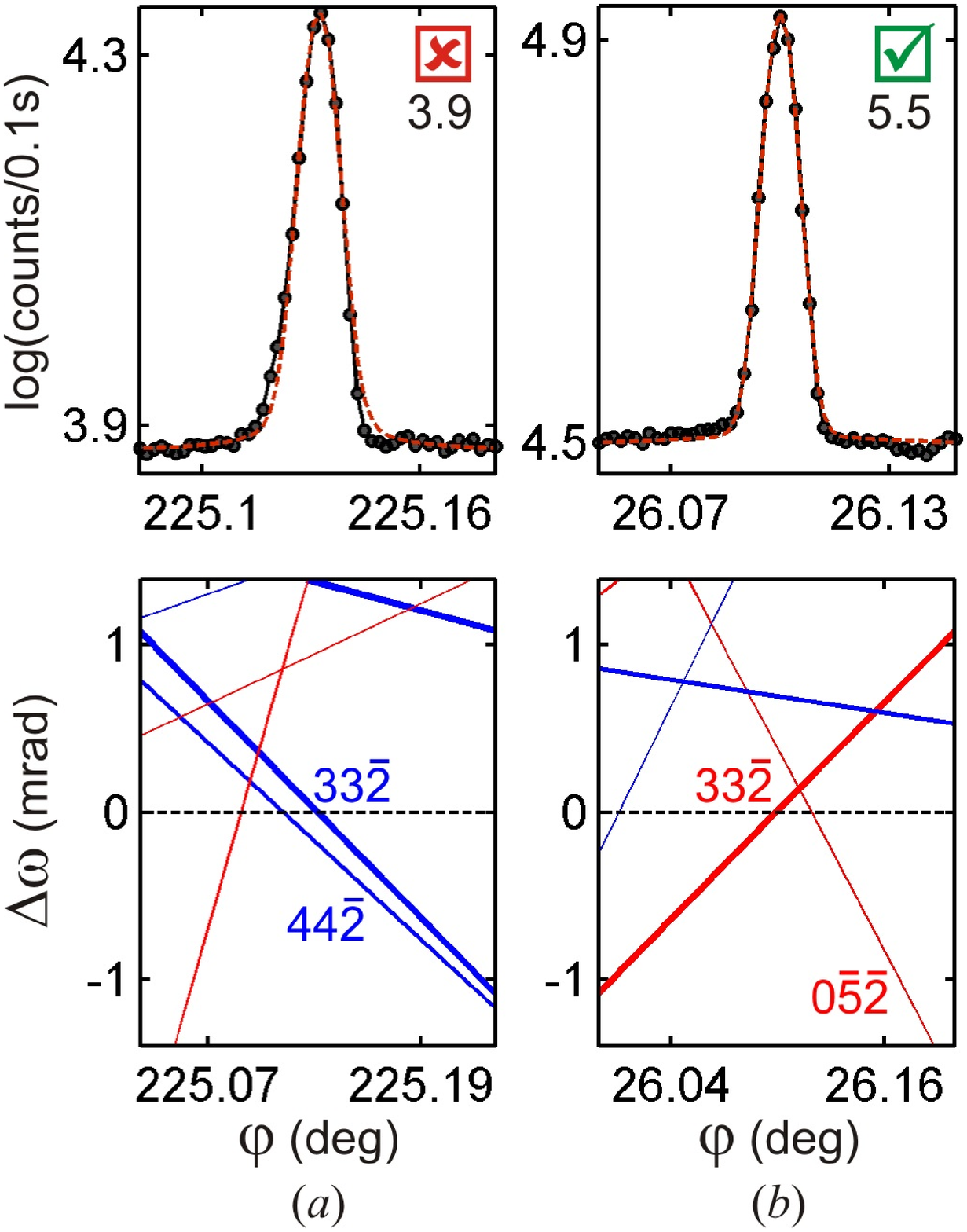}\\
  \caption{(a) Out-in and (b) in-out experimental profiles of a MD case. The H$|$L asymmetry in (a) is caused by the nearby $44\bar{2}$ BC line. $R_a$ values are shown below the true/false checkbox for compatible asymmetry with the NH3 model. Horizontal dashed line stands for $26\bar{1}$ BC line.}
\label{fig:peak332br}
\end{figure}

Being able to discriminate between model structures with subtle differences is the actual challenge to phase measurements. Detecting small shifts in the triplet phases imply in working with nearly symmetrical profiles whose asymmetric character can be influenced by nearby MD cases. Therefore, identification of isolated MDs is a crucial step in testing the compatibility between structure models and profile asymmetries. Coincidental BC lines of comparable strength ($W$ value) crossing the primary BC line at close positions can compromise the asymmetry analysis, as shown for example in Fig.~\ref{fig:peak332br}. When the instrumentation allows the measuments of both out-in and in-out excitation geometries, as in a complete Renninger scan (\S \ref{ap:azscan}), both profiles must present opposite asymmetries. Otherwise, only the one with isolated BC line or with very weak neighbors can be used, as in Fig.~\ref{fig:peak332br}(b).

More refined models are obtained by taking into account small variations in ionic charges. To investigate the polarization of hydrogen bonds, the atomic scattering factors for the amine group are written as $f_{{\rm N}^{3x-}} = (1-x)f_{\rm N}+xf_{{\rm N}^{3-}}$ for the nitrogen and $f_{{\rm H}^{x+}} = (1-x)f_{\rm H}$ for the hydrogens. $x=0$ and $x=1$ are the two extreme situations represented in the NH3 and N$3e$ models, respectively. Phase measurements agree with theoretical phases for $x=0$. However, by slightly changing $x$ we can have a more accurate idea of how susceptible the phases actually are to electron charge distribution at the amine group.

\begin{figure}
\includegraphics[width=2.7in]{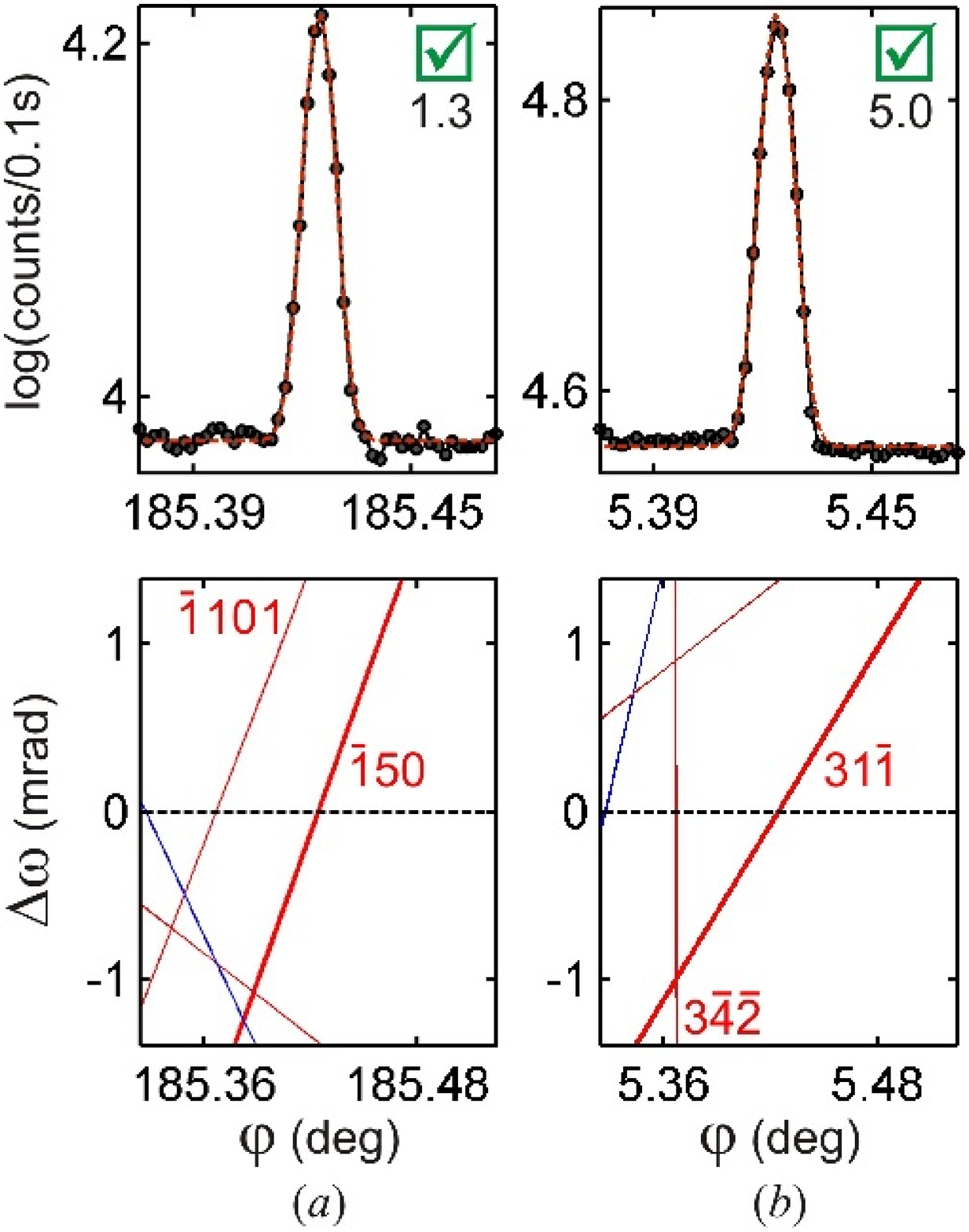}\\
  \caption{Experimental profiles and respective $\omega:\varphi$ graphs of most susceptible cases for polarization of hydrogen bonds. $R_a$ values are shown below the true/false checkbox for compatible asymmetry with the NH3 model. Horizontal dashed lines stand for $26\bar{1}$ BC line.}
\label{fig:mod7profiles}
\end{figure}

For $x=0.1$, shifts in triplet phases of about $\Delta\Psi=\pm4^\circ$ would be enough to invert the line profile asymmetry of a few MD peaks of reasonable amplitudes $W>5$\%; they are indicated as the most susceptible cases in Table~\ref{tab:wlist1} (\S\,\ref{ap:tplist}). Experimentally we are limited to the peaks with isolated BC line and reliable value of asymmetry ($|R_a|\geq1$) that are shown in Fig.~\ref{fig:mod7profiles}. Their asymmetries are consistent to the NH3 model where $x<0.1$, which means that hydrogens in the amine group are practically neutral atoms with effective ionic charges smaller than $+0.1e$.

Compatibility of other models have been also verified. For instance, a model with one electron removed from the nitrogen and shared between the oxygens. Atomic scattering factors are for nitrogen ${\rm N}^+$ and oxygen ${\rm O^{0.5-}}$ ions, while all other atoms are neutral. When compared to the NH3 model, the MD peaks that could present inversion of asymmetry are exactly the same ones previously analyzed in Fig.~\ref{fig:mod7profiles}. Then, there is not evidence to sustain that the electron from the amine group is evenly shared between the two oxygens of the carboxylate group.

\begin{figure*}
\includegraphics[width=5.4in]{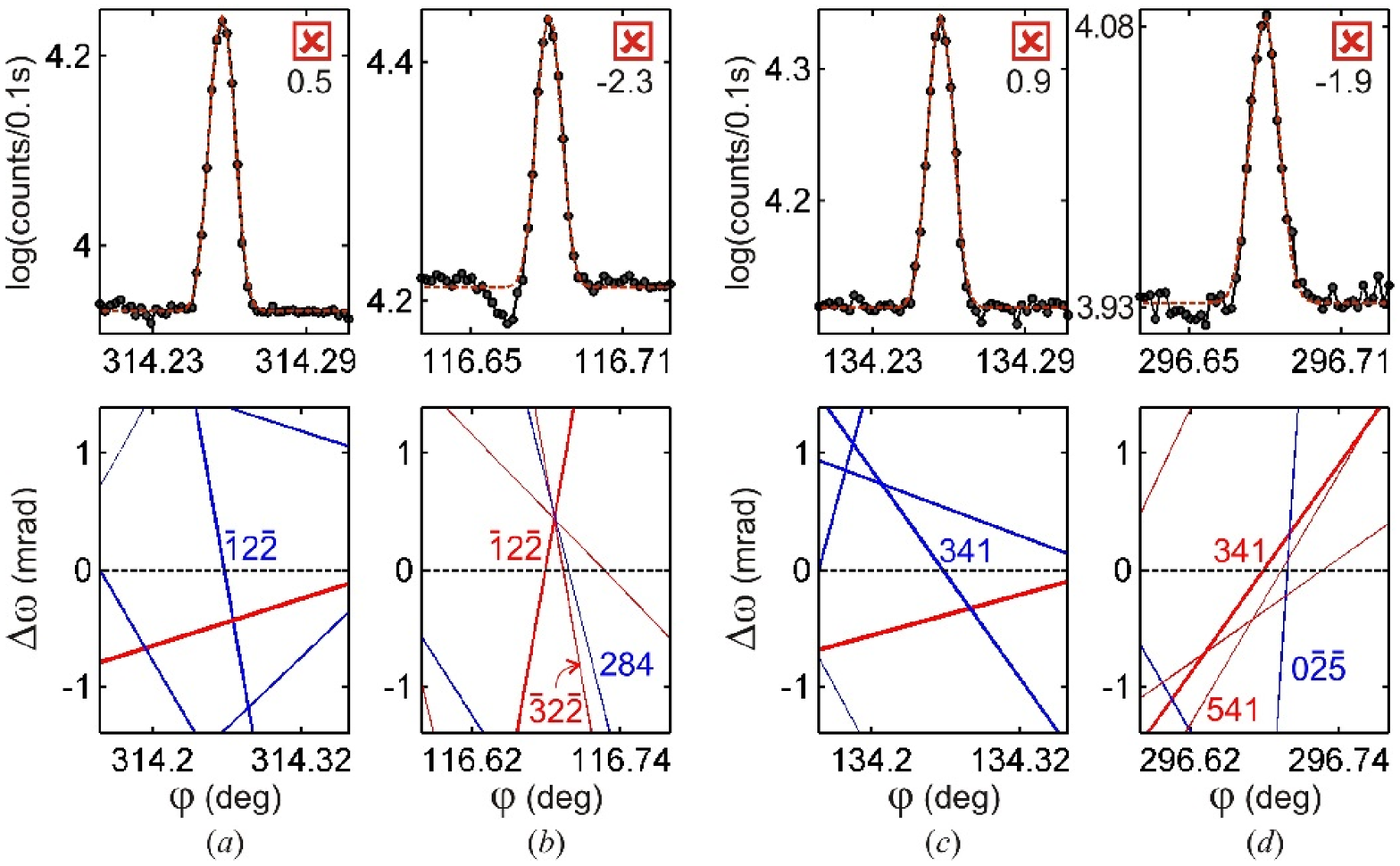}\\
  \caption{Experimental profiles and respective $\omega:\varphi$ graphs of most susceptible cases to a zwitterion. $R_a$ values are shown below the true/false checkbox for compatible asymmetry with the NH3 model. Horizontal dashed lines stand for $26\bar{1}$ BC line.} \label{fig:mod6profiles}
\end{figure*}

A zwitterion model where the electron from the ${\rm N}^+$ ion is placed at the nearest oxygen ${\rm O}^-$ ion, as indicated in Fig.~\ref{fig:xrs} (inset), seems to be compatible to the data. In comparison to the NH3 model, the discrepancies are listed in Table~\ref{tab:wlist4} and the MD peaks of this list that could be measured are shown in Fig.~\ref{fig:mod6profiles}. The phase shifts are very small and can affect only MD cases with $\Psi$ very close to $\pm90^\circ$, whose asymmetric character are difficult to identify. Although, the four profiles agree with the zwitterion model, the most reliable profile is the one in Fig.~\ref{fig:mod6profiles}(c) where there are no nearby BC lines and the asymmetric parameter value $R_a=0.9$ is close to the detectability limit of asymmetry established in Eq.~(\ref{eq:R}).

\begin{table}
\caption{MD cases where $\cos(\Psi)\cos(\Psi^{\prime\prime})<0$ for D-alanine NH3 ($\Psi$) and zwitterion ($\Psi^{\prime\prime}$) models. Secondary $H$ reflections diffracting at azimuth $\varphi_{\rm oi}$ (out-in) and $\varphi_{\rm io}$ (in-out). Primary reflection $G=26\bar{1}$. X-rays of 10\,keV.}\label{tab:wlist4}
\begin{center}
\begin{tabular}{crrcrr}
 \hline\hline
  $H$ & $\Psi\,(^\circ)$ & $\Psi^{\prime\prime}\,(^\circ)$ & $W$(\%) & $\varphi_{\rm oi}\,(^\circ)$ & $\varphi_{\rm io}\,(^\circ)$ \\
\hline
$\bar{1} 2 \bar{2}$ & $-88.8$ & $-90.7$ & $18$ & $314.258$ & $116.679$ \\
$3 4 1$ & $-88.8$ & $-90.7$ & $18$ & $134.258$ & $296.679$ \\
$\bar{4} 7 1$ & $91.3$ & $89.9$ & $8$ & $47.780$ & $157.278$ \\
$6 \bar{1} \bar{2}$ & $91.3$ & $89.9$ & $8$ & $227.780$ & $337.278$ \\
$\bar{1} 2 \bar{3}$ & $90.9$ & $88.6$ & $8$ & $310.281$ & $100.009$ \\
$3 4 2$ & $90.9$ & $88.6$ & $8$ & $130.281$ & $280.009$ \\
$3 \bar{4} \bar{2}$ & $90.9$ & $89.0$ & $6$ & $232.542$ & $5.367$ \\
$\bar{1} 10 1$ & $90.9$ & $89.0$ & $6$ & $52.542$ & $185.367$ \\
\hline \hline
\end{tabular}
\end{center}
\end{table}

\subsection{Radiation damage}\label{ss:rdamage}

The possibility of studying radiation damage on hydrogen bonds arise due to the high sensitivity of triplet phases to the presence of these bonds. Assuming the zwitterion model (${\rm NH}_3^+$-${\rm C}_2{\rm H}_4$-${\rm COO}^-$) with $x$ as the occupancy of H sites at N-H$\cdots$O bonds, the MD case with $H=\bar{1}50$, Fig.~\ref{fig:mod7profiles}(a), has triplet phase $\Psi=-87.4^\circ$ for $x=1$ and $\Psi=-93.6^\circ$ for $x=3/4$ when calculated for X-rays of 8\,keV. These phase values mean that phase measurements can detect 1 missing H atom on every 4 bonds or, equivalently, an average of 3 broken bonds per unit cell [Fig.~\ref{fig:vdWaals}(c)].

\begin{figure*}
\includegraphics[width=5.4in]{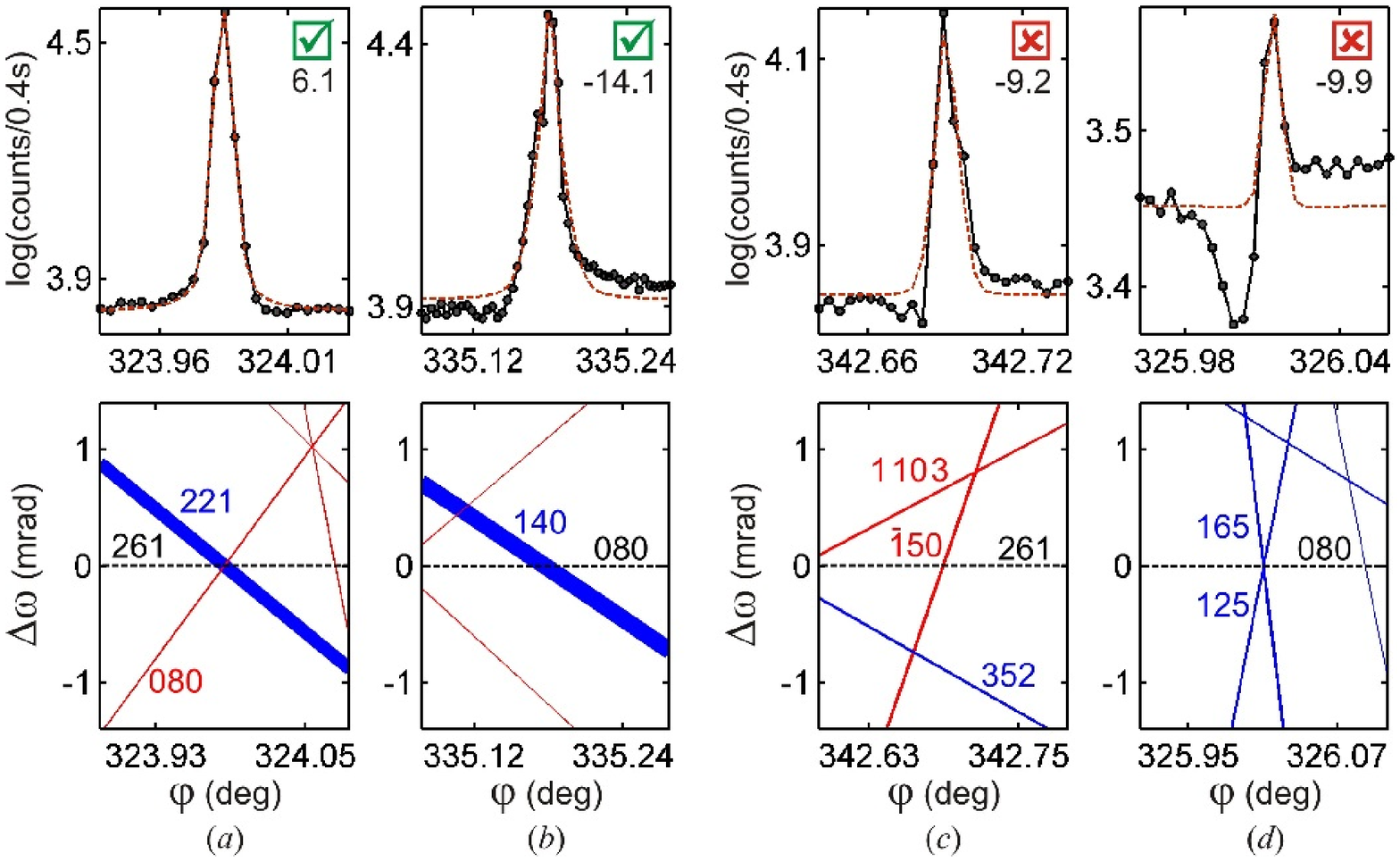}\\
\caption{Experimental profiles and respective $\omega:\varphi$ graphs of a few MD peaks measured using multi-axis goniometry for primary (a,c) 261 and (b,d) 080. $R_a$ values are shown below the true/false checkbox for compatible asymmetry with the NH3 model.}
\label{fig:srsdata}
\end{figure*}

Direct radiation damage on hydrogen bonds can be caused by Compton scattering, whose cross section for H atoms is $0.5748\times10^{-22}\,{\rm mm}^2$. To have one H$^+$ on every four H atoms within a time scale of 10 hours---typical single-crystal experiment---, the required beam flux is $1.2\times10^{17}$\,ph/mm$^2$/s, a too high value for the today's synchrotron sources. However, broken hydrogen bonds as secondary damage caused by collision of any ejected electrons from other atoms demand a much lower flux. The ionization cross section for the entire unit cell of D-alanine is $3962\times10^{-22}\,{\rm mm}^2$ when taking into account Compton and photoelectric processes (see \S\ref{ap:tplist}). Then, 3 ionizations per unit cell in a time period of 10 hours require a flux of $2.1\times10^{14}$\,ph/mm$^2$/s. On a beam size of $50\times200\,\mu{\rm m}^2$, this flux corresponds to an intensity of $2.1\times10^{12}$\,ph/s, well below the direct beam intensity of $10^{13}$\,ph/s available at the I16 beamline of the Diamond Light Source. The visible damage observed after a few seconds of exposure to such direct beam, Fig.~\ref{fig:fullxrs} (inset), can be understandable if each ejected electron is capable of destroying not only one but many hydrogen bonds. When the hydrogen electron is ejected either by Compton or electron collision, the H$^+$ is repealed by the N$^+$ ion preventing any fast mechanism of electron-hole pair recombination to repair the missing bond. Formation of H$_2$ gas is what has been reported instead \cite{am10}.

With the multi-axis goniometry of beamline I16, short azimuthal scans could be perfomed on a few MD cases, including cases on other primary reflections. Most of the profiles agree with the zwitterion model, as those in Figs.~\ref{fig:srsdata}(a,b). But, there were two exception that are shown in Figs.~\ref{fig:srsdata}(c,d). The MD with primary 261 and secondary $\bar{1}50$ has triplet phase very susceptible to the presence of hydrogens bonds as previously discussed on the above paragraphs. Its asymmetry seen in Fig.~\ref{fig:srsdata}(c) is clearly of the L$|$H type, opposite to the one seen in Fig.~\ref{fig:mod7profiles}(a), indicating a phase shift towards a final value of   $\Psi<-90^\circ$. This shift can be explained on basis of radiation damage when more than 25\% of the intermolecular bonds have been broken during data acquisition. The direct beam was attenuated enough to prevent only damages that could be perceived under an optical microscope after hours of exposure. Theoretical ionization rate for such high intensity beam and observed phase shift are in agreement. Nevertheless, both results (theoretical and experimental) should be taken just as evidences suggesting phase measurements as a feasible method to quatify radiation damage at atomic level on biological single crystals. Further investigations under more controlled conditions of flux and time of exposure are still needed to delimit adequate instrumentation and procedures for this type of study.

Profile asymmetries with primary 080 are not susceptible to the subtle variations of model structures discussed in this work. All MD cases for this primary should present asymmetries according to the zwitterion model. It allows us to search for MD cases that are exception to the asymmetry rule in Fig.~\ref{fig:asycrit}. Although the 080 reflection diffracts under Laue-transmission geometry, i.e. the incident and reflected beams are not on the same side of the (130) crystal surface, the only rule exception we found, shown in Fig.~\ref{fig:srsdata}(d), has poor sensitivity to the triplet phase due to polarization suppression of the second-order term of dynamical coupling responsible for the phase information \cite{tho98,yps00,slm02}. A situation that occurs when the $H$ reflection has Bragg angle close to 45$^\circ$, as the 125 reflection (Bragg angle of $43.9^\circ$), and diffracting in $\pi$-polarization, i.e. its BC line appears nearly vertical in the $\omega\!:\!\varphi$ graph for the used beamline setup. Another situation compromising direct phase evaluation occurs for MD cases with very weak \textit{Umweganregung} and strong \textit{Aufhellung} components \cite{wec97,ros99}. Such cases are easily avoided when a very weak reflection can be picked up as primary reflection. Otherwise, MD cases with very weak or polarization suppressed $G-H$ coupling reflections are those with poor reliability for phase measurements.

\section{Conclusions}

The greatest achievement of this work is to have demonstrated in practice the full potential of phase measurements applied to current trends in crystallography. Hydrogen bonds were easily detected, a maximum value attributed to their effective polarization, model structures with subtle variations in ionic charges discriminated, relevant informations for molecular dynamics studies of this amino acid crystal obtained, and insights on quantitative analysis of radiation damage discussed on theoretical and experimental grounds. Besides a model where all atoms are neutral, the only other model that can agree with the whole data set of MD profiles asymmetries is the zwitterion model where an electron from the nitrogen orbitals goes to the nearest oxygen. In this case, the oxygens in carboxylate group have different ionic charges and the intermolecular forces stabilizing the D-alanine crystal are also van der Waals forces between ${\rm N}^+\!\rightarrow{\rm O}^-$ electrical dipoles. Phase sensitivity to the average number of hydrogen bonds per unit cell and experiments using high flux synchrotron radiation point towards a damage mechanism where most of bond cleavage are caused by photoelectron collisions. A whole package of experimental and data analysis procedures are given and explained in details, allowing immediate use of phase measurement on a wide range of studies. The only requirements are crystals of good quality capable of undergo dynamical diffraction and availability of suitable structure models to each especific features in the crystal electron density to be investigated.

\begin{acknowledgments}
Acknowledgments are due to the Brazilian funding agencies CNPq (Grant Nos. 306982/2012-9 and 452031/20150), FAPESP (Grant Nos. 2012/01367-2, 12/15858-8, 14/08819-1, 14/21284-0, and 16/11812-4), Diamond Light Source (proposal MT11922), Brazilian Synchrotron Light Source (proposals 17063, 18011, and 19018). We also thank Professor Lisandro P. Cardoso, Dr. Steven Collins, and Dr. Jos\'e Brand\~ao-Neto for helpful discussions.
\end{acknowledgments}

\appendix
\section{Supplemental Information}

\subsection{Physical alignment of primary reflection by using the rotating crystal method}\label{ap:cara}

\begin{figure}
\includegraphics[width=2.7in]{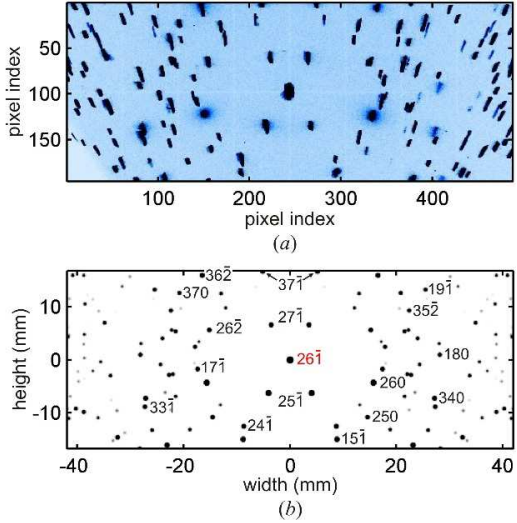}\\
\caption{(a) Diffraction spots on detector area collected in a $360^\circ$ spin of the sample around diffraction vector of reflection $26\bar{1}$. Sample-detector distance is 74.7\,mm. X-rays of 10\,keV, $\sigma$ polarization. (b) Indexing of diffraction spots by simulation of the rotating crystal method \cite{rcm}. Area detector width at horizontal direction.} \label{fig:rcm}
\end{figure}

With an area detector (the pilatus 100K in our case) attached to the $2\theta$ arm of the single crystal diffractometer, 2D array indices $m_cn_c$ of the reference pixel receiving most of the direct beam ($2\theta=0$) are identified. After fixing the sample in a goniometer head that has arcs for orientation correction, placing the area detector at a distance $D$ from the sample, and moving the detector arm to a desired $2\theta$ value, the $\varphi$ axis is spun by $360^\circ$ while the detector acquires images at the rate of one frame per degree of rotation. Diffraction spots at pixels of indices $mn$ correspond to reflections with scattering angles $2\theta_{mn}=\cos^{-1}(\hat{R}\cdot\hat{z})$ where $\bfm{R} = |\bfm{R}|\hat{R} = \bfm{R}_c + \bfm{r}_d$, $\bfm{R}_c = D[\sin(2\theta)\hat{x} +\cos(2\theta)\hat{z}]$, $\bfm{r}_d=-(n-n_c)p\hat{x}_d + (m-m_c)p\hat{y}_d$, $\hat{x}_d =\cos(2\theta)\hat{x}-\sin(2\theta)\hat{z}$, and $\hat{y}_d = \hat{y}$. In the lab frame, $\hat{z}$ is along the direct beam, $\hat{x}$ is vertical, $\hat{y}$ is horizontal. Pixel size for the used detector is $p=0.172$\,mm.

A diffraction spot with scattering angle close to that of the desired primary reflection is selected and aligned to the $\varphi$ axis. By using as input pixel indices $mn$ and azimuth $\varphi_{mn}$ of a spot, a script suitable to the used diffractometer was written to provide the values by which each arc has to be corrected. A new spin of the rotation axis is then carried out to confirm that the aligned reflection is in fact the desired one, as shown in Fig.~\ref{fig:rcm} for the $26\bar{1}$ reflection.

\subsection{Azimuthal scanning with a single goniometer axis}\label{ap:azscan}

\begin{figure}
\includegraphics[width=2.7in]{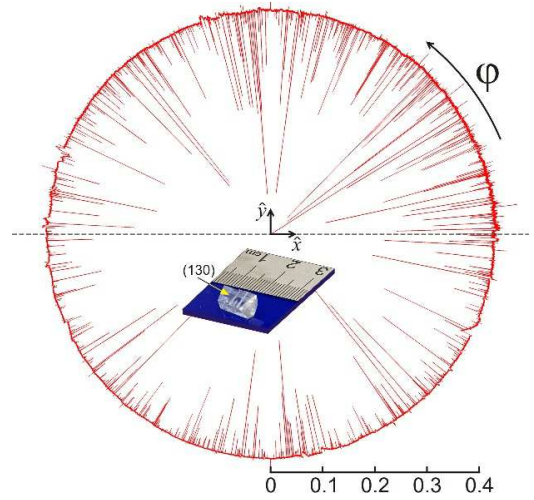}\\
  \caption{Complete Renninger scan of reflection $26\bar{1}$ in polar plot: $x=-\log(I/I_{max})\cos\varphi$ and $y=-\log(I/I_{max})\sin\varphi$. $I_{max}=1.8\times10^5$\,counts/0.1s. X-rays of 10\,keV, $\sigma$ polarization. Reference direction ($\varphi=0$): $c$ axis in the incident plane pointing downstream. Sense of sample rotation: clockwise with the diffraction vector pointing to the observer. Inset: used sample showing streaks on face (130) caused by the direct beam.} \label{fig:fullxrs}
\end{figure}

The full Renninger scan of the $26\bar{1}$ reflection shown in Fig.~\ref{fig:fullxrs} is composed of 72 uninterrupted $\varphi$ scans. At the first point of each scan, the primary reflection has been centered at its rocking-curve (FWHM of 0.07\,mrad). The maximum drift of the rocking-curves center as a function of $\varphi$ was 0.2\,mrad. Only in Fig.~\ref{fig:fullxrs}, smooth variations of the baseline intensity have been flattened for the sake of visualization of the $180^\circ$ symmetry of the data, although peak intensities are different since the (130) entrance surface normal direction is not the one aligned to the rotation axis.

\subsection{Calculation codes for triplet phases and ionization cross sections}\label{ap:tplist}

\begin{table}
\caption{Partial list of MD cases in which $\cos(\Psi)\cos(\Psi^\prime)<0$ regarding the NH3 ($\Psi$) and N$3e$ ($\Psi^\prime$) structure models of D-alanine. Secondary $H$ reflections diffracting at azimuth $\varphi_{\rm oi}$ (out-in) and $\varphi_{\rm io}$ (in-out). Primary reflection $G=26\bar{1}$. X-rays of 10\,keV. $^\dag$Susceptible cases to polarization of hydrogen bonds.}\label{tab:wlist1}
\begin{center}
\scriptsize{
\begin{tabular}{crrcrr}
 \hline\hline
  $H$ & $\Psi\,(^\circ)$ & $\Psi^\prime\,(^\circ)$ & $W$(\%) & $\varphi_{\rm oi}\,(^\circ)$ & $\varphi_{\rm io}\,(^\circ)$ \\
\hline
$0 4 0$ & $-98.1$ & $-60.4$ & $100$ & $12.199$ & $212.730$ \\
$2 2 \bar{1}$ & $-98.1$ & $-60.4$ & $100$ & $192.199$ & $32.730$ \\
$\bar{1} 3 1$ & $72.9$ & $111.1$ & $31$ & $45.132$ & $206.099$ \\
$3 3 \bar{2}$ & $72.9$ & $111.1$ & $31$ & $225.132$ & $26.099$ \\
$1 1 0$ & $-102.3$ & $-65.4$ & $29$ & $130.286$ & $357.040$ \\
$1 5 \bar{1}$ & $-102.3$ & $-65.4$ & $29$ & $310.286$ & $177.040$ \\
$2 2 0$ & $75.0$ & $112.3$ & $28$ & $143.710$ & $343.615$ \\
$0 4 \bar{1}$ & $75.0$ & $112.3$ & $28$ & $323.710$ & $163.615$ \\
$0 1 2$ & $-116.0$ & $-79.5$ & $25$ & $98.710$ & $246.456$ \\
$2 5 \bar{3}$ & $-116.0$ & $-79.5$ & $25$ & $278.710$ & $66.456$ \\
$1 \bar{2} 2$ & $79.9$ & $112.8$ & $23$ & $151.438$ & $279.499$ \\
$1 8 \bar{3}$ & $79.9$ & $112.8$ & $23$ & $331.438$ & $99.499$ \\
$2 \bar{2} 0$ & $75.3$ & $112.5$ & $23$ & $188.775$ & $346.556$ \\
$0 8 \bar{1}$ & $75.3$ & $112.5$ & $23$ & $8.775$ & $166.556$ \\
$1 \bar{2} \bar{2}$ & $79.0$ & $118.0$ & $23$ & $249.286$ & $43.466$ \\
$1 8 1$ & $79.0$ & $118.0$ & $23$ & $69.286$ & $223.466$ \\
$\bar{1} 5 0^\dag$ & $89.8$ & $125.5$ & $19$ & $15.689$ & $185.424$ \\
$3 1 \bar{1}^\dag$ & $89.8$ & $125.5$ & $19$ & $195.689$ & $5.424$ \\
$2 \bar{6} 0$ & $75.5$ & $112.7$ & $19$ & $221.416$ & $331.212$ \\
$0\, 12\, \bar{1}$ & $75.5$ & $112.7$ & $19$ & $41.416$ & $151.212$ \\
$0 2 0$ & $79.1$ & $118.8$ & $18$ & $1.584$ & $223.346$ \\
$2 4 \bar{1}$ & $79.1$ & $118.8$ & $18$ & $181.584$ & $43.346$ \\
$\bar{1} 1 \bar{3}^\dag$ & $89.7$ & $130.9$ & $17$ & $306.476$ & $93.272$ \\
$3 5 2^\dag$ & $89.7$ & $130.9$ & $17$ & $126.476$ & $273.272$ \\
$2 2 3$ & $73.1$ & $107.5$ & $17$ & $126.720$ & $267.779$ \\
$0 4 \bar{4}$ & $73.1$ & $107.5$ & $17$ & $306.720$ & $87.779$ \\
$2 2 1$ & $-111.9$ & $-75.2$ & $17$ & $126.354$ & $304.583$ \\
$0 4 \bar{2}$ & $-111.9$ & $-75.2$ & $17$ & $306.354$ & $124.583$ \\
$2 2 \bar{3}$ & $-102.6$ & $-62.6$ & $16$ & $256.833$ & $54.872$ \\
$0 4 2$ & $-102.6$ & $-62.6$ & $16$ & $76.833$ & $234.872$ \\
$\bar{1} \bar{5} \bar{1}$ & $78.0$ & $114.8$ & $14$ & $333.637$ & $351.023$ \\
$3\, 11\, 0$ & $78.0$ & $114.8$ & $14$ & $153.637$ & $171.023$ \\
$2 4 1$ & $71.4$ & $106.1$ & $13$ & $108.155$ & $286.344$ \\
$0 2 \bar{2}$ & $71.4$ & $106.1$ & $13$ & $288.155$ & $106.344$ \\
$\bar{3} 9 \bar{1}$ & $62.4$ & $100.2$ & $13$ & $23.392$ & $146.181$ \\
$5 \bar{3} 0$ & $62.4$ & $100.2$ & $13$ & $203.392$ & $326.181$ \\
$0 \bar{4} 3$ & $71.8$ & $109.0$ & $13$ & $178.286$ & $228.492$ \\
$2\, 10\, \bar{4}$ & $71.8$ & $109.0$ & $13$ & $358.286$ & $48.492$ \\
$2 \bar{2} \bar{3}$ & $73.2$ & $107.7$ & $11$ & $251.254$ & $34.538$ \\
$3 3 \bar{3}$ & $-115.7$ & $-71.6$ & $11$ & $250.307$ & $35.484$ \\
$3 0 3^\dag$ & $88.9$ & $125.2$ & $10$ & $150.154$ & $280.783$ \\
$\bar{1} 6 \bar{4}^\dag$ & $88.9$ & $125.2$ & $10$ & $330.154$ & $100.783$ \\
$\bar{2} \bar{1} 0$ & $-111.1$ & $-73.8$ & $9$ & $30.734$ & $114.471$ \\
$1 0 \bar{5}^\dag$ & $88.9$ & $125.3$ & $8$ & $288.312$ & $52.335$ \\
$1 6 4^\dag$ & $88.9$ & $125.3$ & $8$ & $108.312$ & $232.335$ \\
$0 8 0$ & $68.5$ & $105.5$ & $7$ & $32.730$ & $192.199$ \\
$2 \bar{9} \bar{3}^\dag$ & $-91.1$ & $-54.0$ & $7$ & $272.849$ & $351.002$ \\
$0\, 15\, 2^\dag$ & $-91.1$ & $-54.0$ & $7$ & $92.849$ & $171.002$ \\
$\bar{3} 9 0$ & $-111.3$ & $-72.4$ & $6$ & $34.859$ & $157.769$ \\
$3 7 \bar{5}$ & $64.5$ & $103.9$ & $6$ & $312.753$ & $30.450$ \\
\hline \hline
\end{tabular}
}
\end{center}
\end{table}

Structure factors taking into account non-resonant and resonant terms of the atomic scattering factors were effectively calculated by routine \texttt{sfactor.m}. It lists the structure factors used for comparison of phase values, as in Fig.~\ref{fig:comparators} and Tables~\ref{tab:phases}, \ref{tab:wlist4}, and \ref{tab:wlist1}. This routine can be found in open codes at the internet \cite{slm16}, as well as the routines \texttt{sgcompton.m} and \texttt{fpfpp.m} used for calculating the Compton and photoelectric absorption cross sections, respectively.

%\bibliography{manuscriptArXiv_dalanine}

\end{document}